\documentclass[12pt]{article}
\usepackage{mathrsfs, amssymb, amsmath, amsfonts, txfonts, latexsym, graphicx,dsfont}  
\usepackage[utf8]{inputenc}
\usepackage{soul}
\usepackage{physics}
\usepackage{accents}
\usepackage{ulem}
\usepackage{cite}
\usepackage[top=2cm, bottom=2cm, right=2cm, left=2cm]{geometry}

\newcommand{\hg}{\hat{g}}

\newcommand{\CQCS}{\mathcal{C}_{\mathrm{QCS}}}

\makeatletter
\DeclareRobustCommand{\loplus}{\mathbin{\mathpalette\dog@lsemi{+}}}

\usepackage{float}
\usepackage{mathtools}
\newcommand{\defeq}{\vcentcolon=}

\newcommand{\stintb}[2]{\int_{#1}\dd \Sigma_{#2}\,}
\newcommand{\dog@lsemi}[2]{\dog@semi{#1}{#2}{270,90}}
\newcommand{\dog@semi}[3]{%
  \begingroup
  \sbox\z@{$\m@th#1#2$}%
  \setlength{\unitlength}{\dimexpr\ht\z@+\dp\z@\relax}%
  \makebox[\wd\z@]{\raisebox{-\dp\z@}{%
    \begin{picture}(1,1)
    \linethickness{\variable@rule{#1}}
    \roundcap
    \put(0.5,0.5){\makebox(0,0){\raisebox{\dp\z@}{$\m@th#1#2$}}}
    \put(0.5,0.5){\arc[#3]{0.5}}
    \end{picture}%
  }}%
  \endgroup
}
\newcommand{\variable@rule}[1]{%
  \fontdimen8  
  \ifx#1\displaystyle\textfont3\else
    \ifx#1\textstyle\textfont3\else
      \ifx#1\scriptstyle\scriptfont3\else
        \scriptscriptfont3\relax
  \fi\fi\fi
}
\makeatother
\usepackage{tikz-cd}

\usetikzlibrary{arrows.meta,calc}

\usepackage{pict2e}
\usepackage{diagbox}

\newcommand{\beq}{\begin{eqnarray}}
\newcommand{\eeq}{\end{eqnarray}}
\newcommand{\beqn}{\begin{eqnarray}}
\newcommand{\eeqn}{\end{eqnarray}}

\newcommand{\RR}{\mathbb{R}}

\newcommand{\spl}[1]{\mathrm{SL}(#1,\mathbb{R})}
\newcommand{\spla}[1]{\mathfrak{sl}\left(#1,\mathbb{R}\right)}

\usepackage{pict2e}

\newcommand{\chkM}{{\color{red} \,\checkmark\kern-5pt{}_{M}}}

\newcommand{\ee}{\end{equation}}
\newcommand{\bea}{\begin{eqnarray}}
\newcommand{\eea}{\end{eqnarray}}

\usepackage{BOONDOX-cal}

\usepackage{scalerel}
\usepackage{stackengine,wasysym}

\newcommand\reallywidetilde[1]{\ThisStyle{%
  \setbox0=\hbox{$\SavedStyle#1$}%
  \stackengine{-.1\LMpt}{$\SavedStyle#1$}{%
    \stretchto{\scaleto{\SavedStyle\mkern.2mu\AC}{.5150\wd0}}{.6\ht0}%
  }{O}{c}{F}{T}{S}%
}}

\newcommand{\os}{\,\hat{=}\,}

\newenvironment{Align}{\begin{equation}
\begin{aligned}}
{\end{aligned}
\end{equation}}
\newenvironment{Align*}{\begin{equation*}
\begin{aligned}}
{\end{aligned}
\end{equation*}}

\usepackage{color}

\usepackage{layout}
\usepackage{hyperref}

\usepackage{cleveref}

\DeclareFontFamily{OT1}{rsfs}{}
\DeclareFontShape{OT1}{rsfs}{m}{n}{ <-7> rsfs5 <7-10> rsfs7 <10->rsfs10}{} 
\DeclareMathAlphabet{\mycal}{OT1}{rsfs}{m}{n}

\DeclareFontFamily{U}{MnSymbolC}{}
\DeclareSymbolFont{MnSyC}{U}{MnSymbolC}{m}{n}
\DeclareFontShape{U}{MnSymbolC}{m}{n}{
    <-6>  MnSymbolC5
   <6-7>  MnSymbolC6
   <7-8>  MnSymbolC7
   <8-9>  MnSymbolC8
   <9-10> MnSymbolC9
  <10-12> MnSymbolC10
  <12->   MnSymbolC12}{}
\DeclareMathSymbol{\intprod}{\mathbin}{MnSyC}{'270}
 \usepackage{authblk}

\title{
From the Corner Proposal to the Area Law}
\author[1,2]{Jerzy Kowalski-Glikman\footnote{Email: jerzy.kowalski-glikman@uwr.pl.edu}}
\author[1]{Ludovic Varrin\footnote{Email:  ludovic.varrin@ncbj.gov.pl}}
\affil[b]{National Centre for Nuclear Research, Pasteura 7, 02-093 Warsaw, Poland, }
\affil[c]{Faculty of Physics and Astronomy, University of Wroclaw, Pl. Maksa Borna 9, 50-204
Wroclaw}

\date{}
\begin{document}

\maketitle
\begin{abstract}
    We provide an explicit realization of the Corner Proposal for Quantum Gravity in the case of spherically symmetric spacetimes in four dimensions, or equivalently, two-dimensional dilaton gravity. We construct coherent states of the Quantum Corner Symmetry group and compute the entanglement entropy relative to these states. We derive the classical corner charges and relate them to operator expectation values in coherent states. For a subset of coherent states that we call classical states, we find that the entanglement entropy exhibits a leading term proportional to the area, recovering the Bekenstein–Hawking area law in the semiclassical limit. 
\end{abstract}

\vspace{1cm}

For more than a century, the study of symmetries and their realizations has been one of the principal driving forces behind developments in fundamental theoretical physics. It is often remarked that “elementary particle physics is the study of the representations of the Poincaré group” \cite{Wigner:1939cj, Weinberg:1995mt}, underscoring the central role played by the symmetry group of Special Relativity in shaping our understanding of the microworld. In a similar vein—and with only a modest degree of exaggeration—one might say that string theory constitutes the representation theory of the (super-)Virasoro algebra, while Loop Quantum Gravity may be viewed as the representation theory of the group of spacetime diffeomorphisms.

In all these frameworks, representation theory furnishes the formal backbone that enables one to impose quantum constraints associated with local spacetime symmetries. This is achieved by systematically reducing the original kinematical Hilbert space to the physical Hilbert space—the space of states annihilated by the relevant constraint operators. For decades, this procedure was regarded as the unique and internally consistent method of realizing gauge symmetries at the quantum level, serving as both a guiding principle and a test of theoretical coherence in attempts to quantize spacetime itself.

In recent years, however, a potential shift in perspective has emerged. It has been recognized \cite{Wald:1993nt, Wald:1999wa, Donnelly:2016auv, Ciambelli:2021vnn, Ciambelli:2022cfr} that the algebra of Noether charges associated with diffeomorphism invariance in general relativity— anchored at the boundary $S$ of a {finite} region (a codimension-2 surface)—forms the so-called Extended Corner Symmetry (ECS) algebra, which is universal and takes the form
\begin{equation}\label{1}
    \mathfrak{ecs} = \mathfrak{diff}(S) \loplus \qty(\spla{2}\loplus \RR^2)^S
\end{equation}

As a consequence of its universality, it is natural to consider that the ECS algebra may play, in quantum gravity, a role analogous to that of the Poincaré algebra—the algebra of Noether charges associated with spacetime symmetries of Minkowski space—in particle physics. This is the essence of the Corner Proposal for Quantum Gravity \cite{Ciambelli:2021vnn,Ciambelli:2022vot,Ciambelli:2022cfr,Ciambelli:2023bmn}, which we understand as a statement that, as the first step towards the construction of full quantum gravity, one should understand the representation theory of the Extended Corner Symmetry. Once the representation theory of the algebra \eqref{1} is understood, we can turn to addressing questions of interest. The first, obvious one would be to see what happens when two regions are glued together to form a single, larger region. Conversely, given the large region, could we split it into two subregions? If so, what would be the entanglement entropy associated with the boundary separating the two subregions? These questions were posed in \cite{Donnelly:2016auv}, and recently we were able to address them in the context of the Corner Proposal \cite{Ciambelli:2024qgi, Ciambelli:2025ztm}. The answer we have obtained is, in a sense, purely quantum: the quantum operators and states do not, a priori, correspond to any (semi)classical spacetime or metrics within them. In particular, there is a priori no way to associate the entanglement entropy with the area of the entangling surface, simply because in quantum gravity the concept of area makes sense only (semi)classically.

Since we are in possession of the representation theory, the entanglement entropy, which is abstractly defined in terms of a functional on the $C^*$ algebra of operators, can now be expressed as a sum over expectation values,  $S_E=\sum_I \bra{I}\rho_R \log\rho_R\ket{I}$, where $\rho_R$ is the reduced density matrix and $\ket{I}$ is an appropriate set of states. In order to relate this entanglement entropy with the area, we must use states that admit a semiclassical interpretation. The natural choice is to use coherent states, which by construction are “as close to classical as possible.” To compare the resulting entanglement entropy with the area, one has to identify a quantum operator that serves as the counterpart of the classical geometrical notion of area. There is a remarkable feature of the Corner Proposal that such an association can indeed be made. Indeed, the Noether charges—the phase-space functions forming, through their Poisson brackets, the ECS algebra \eqref{1}—are in one-to-one correspondence with quantum operators forming the quantum counterpart of the ECS algebra through their commutators. In this way, given semiclassical states, we can associate the expectation values of these operators with the geometric data carried by the classical Noether charges of the ECS. It is the aim of this letter to make this idea explicit in the context of spherically symmetric spacetimes in four dimensions—or, equivalently, in two-dimensional gravity coupled to a dilaton field.
\newline

In the recent papers \cite{Ciambelli:2024qgi, Ciambelli:2025ztm} we started this research program by investigating the representation theory of ECS in two spacetime dimensions. In this case, the corner $S$ reduces to a point, and \eqref{1} simplifies considerably due to the disappearance of the diffeomorphism part
\begin{equation}\label{2}
    \mathfrak{ecs}_2 = \qty(\spla{2}\loplus \RR^2)
\end{equation}
As explained in \cite{Ciambelli:2024qgi,Ciambelli:2025ztm} quantization of the theory is then realized on projective representations of the ECS$_2$ group, which correspond to standard representations of the maximally centrally extended group: the quantum corner symmetry (QCS) group
 \begin{equation}
     QCS = \reallywidetilde{\mathrm{SL}}\qty(2,\RR) \ltimes H_3,
 \end{equation}
 where $H_3$ is the three-dimensional Heisenberg group. Those representations can be found through the orbit method \cite{Neri:2025fsh}, or Mackey's theory of induced representations \cite{Varrin:2024sxe}. They are given by tensor products of the representations of 
$\reallywidetilde{\mathrm{SL}}\qty(2,\RR)$ with those of $H_{3}$. The special linear elements act through the Weil representation on the Heisenberg side. In the present work, we will use the positive discrete series representations. In order to describe them, we introduce the ladder basis of the QCS: $\qty{L_0,L_\pm,P_\pm,C}$ with the following non-vanishing commutator
\begin{Align}
    \qty[L_0,L_\pm] &= \pm L_\pm,\quad \qty[L_-,L_+] = 2 L_0,\\
    \qty[L_\pm,P_\mp] &= \mp P_\pm, \quad \qty[L_0,P_\pm] = \pm \frac12 P_\pm,\\
    &\,\,\: \qty[P_-,P_+] = C.
\end{Align}
The QCS algebra admits two Casimir. The first one is the central generator $C$, and the second one is the  combination
\begin{Align}\label{eq:qcscubiccasimir}
      \mathcal{C}_{\mathrm{QCS}} = \qty(L_0( L_0 + \frac32)- L_- L_+ + \frac{1}{16}) +\frac{1}{2 C}\left(L_- P_+^2 + L_+ P_-^2 -2 L_0 P_- P_+\right)
\end{Align}
The {constant shift} $\tfrac{1}{16}$ is added to ensure that the operator is positive semi-definite in any of the positive discrete series representations. Note that in the limit where the central element goes to zero, the Casimir is dominated by the second term which is the Casimir operator of the $\mathrm{ECS}$.

The positive discrete series representations can then be expressed in the orthonormal Fock basis 
$\bigl\{\,\ket{n,k}\;\bigm|\; n,k \in \mathbb{N}\,\bigr\}$ as
\begin{Align}\label{discreteseriesrepresentation}
    P_- \ket{n,k} &= \sqrt{ ck} \ket{n,k-1},\\
    P_+ \ket{n,k} &= \sqrt{c (k+1)} \ket{n,k+1},\\
    L_0 \ket{n,k} &= \qty(s +\frac{k}{2}+n+\frac{3}{4})\ket{n,k},\\
    L_- \ket{n,k} &= \sqrt{n (n+2 s)}\ket{n-1,k} + \frac12\sqrt{k(k-1)}\ket{n,k-2},\\
    L_+ \ket{n,k} &= \sqrt{(n+1) (2 s +n+1)}\ket{n+1,k}+\frac12\sqrt{(k+1)(k+2)}\ket{n,k+2}.
\end{Align}
The lowest weight "vacuum" state $\ket{\Omega} = \ket{0,0}$ is annihilated by both lowering operators
\begin{equation}\label{9}
    L_- \ket{\Omega} = P_- \ket{\Omega} = 0,
\end{equation}
and the representations are labeled by the parameters $c,s \in \RR_+$ which are related to the QCS Casimirs 
\begin{align}
    C\ket{n,k} &=  c\ket{n,k},\\
    \CQCS \ket{n,k} &= s^2 \ket{n,k}.
\end{align}
{While the above was written for $c>0$, the case $c<0$ can be treated analogously, with the Weil sector realized as a highest-weight representation instead.}

It should be emphasized that the vacuum state \eqref{9}, like all other $\ket{n,k}$ states, carries no intrinsic spacetime-geometrical meaning. At this stage, we are dealing solely with the algebra of operators and their discrete representations, distinguished by the presence of a lowest-weight state. Some states—or particular subclasses among them—may permit a semiclassical interpretation, wherein the expectation values of the operators acquire a correspondence with classical spacetime quantities. In general, however, these states do not represent any spacetime geometry.

We call quantum corner states those states that can be expressed as a linear combination of the Fock basis states. In order to describe a glued corner ---the entangling surface between two subregions--- we perform the gluing procedure first described in \cite{Ciambelli:2024qgi}. We begin with two separate corners, each described by a single state in the QCS representation, and construct their tensor product. The identification of these two corners is then imposed by requiring that the quantum charges take the same values when acting on either the left or the right of the tensor product. Since the underlying algebra is non-Abelian, the strongest condition that can be imposed without annihilating the state altogether is the equality of the charges belonging to the maximally commuting subalgebra. In order to explicitly construct the glued states, we introduce a  basis of (dimensionful) Hermitian operators, which will later play a role in the identification with the corresponding classical charges
\begin{Align}\label{eq:hermitianbasis}
    H &= \hbar\left(L_0 -\frac12 \qty( L_++L_-)\right),\\
    D &= \frac{i \hbar}{2}(L_+ - L_-),\\
    K &= \hbar\left(L_0 + \frac12(L_++L_-)\right),\\
    X &= \sqrt{\frac{\hbar}{2 }}(P_++P_-),\\
    P &= i\sqrt{\frac{\hbar}{2}}(P_+-P_-).
\end{Align}

Since our aim here is to extract semiclassical information from quantum corners, we now turn to the construction of an appropriate class of states that approximate a semiclassical geometry as closely as a quantum state can. In standard quantum mechanics, it is well known that the Glauber coherent states parametrize the classical phase space (the $x$ and $p$) \cite{Glauber:1963tx} (for review see \cite{Rosas-Ortiz:2018jum}). It is possible to define a generalization of this property for arbitrary Lie group G. For a group $G$ with a unitary irreducible representation $D$ on a Hilbert space $\mathcal{H}$, such generalized coherent states are called Perelomov coherent states. They are the “most classical” in the sense that their parameters serve as coordinates on the coadjoint orbits.

In the case of the QCS algebra, the construction of Perelomov coherent states proceeds as follows. We first pick a fiducial state $\ket{\psi_0}\in \mathcal{H}$, which in our case is naturally chosen to be the vacuum $\ket{\Omega}$. The QCS coherent state are then defined as
\begin{equation}\label{eq:qcscoherentstate}
    \ket{\zeta,\alpha} \defeq \mathcal{D}(\alpha)\mathcal{S}(\zeta)\ket{\Omega} = e^{\frac{1}{\sqrt{c}}(\alpha P_+ -\bar{\alpha}P_-)}e^{c_\zeta L_+ - \bar{c}_\zeta L_-}\ket{\Omega} .
\end{equation}
where, denoting $\zeta = re^{i\theta}$
\begin{equation}
    c_\zeta =\tanh^{-1}(r)e^{-i\theta},
\end{equation}
This particular parametrization is a choice of coordinates $\zeta,\alpha \in \mathbb{C}$. Any other parametrization would only alter the phase of the coherent state. 

As shown in \cite{Neri:2025fsh}, for the positive discrete series, the complex $\zeta$ coordinates should lie in the unit disk $\abs{\zeta} \leq 1$. Moreover the coherent state \eqref{eq:qcscoherentstate} decomposes into a product of a $\spl{2}$ Perelomov coherent state  \cite{APerelomov_1977} and a squeezed Glauber coherent state \cite{Stoler:1969tq,Yuen:1976vy}
\begin{align}\label{classcohstate1}
    \ket{\zeta,\alpha} =  \ket{\zeta} \otimes \ket{\alpha_\zeta}
\end{align}

The properties of the coherent states $\ket{\zeta,\alpha}$ have been discussed in details in \cite{Ciambelli:2025ztm}; here we will restrict our attention to the particular class of these coherent states that we call \textit{classical states} defined by
\begin{align}\label{eq:classicalstateparameter}
    c_\zeta = \bar c_\zeta = -s 
\end{align}
In the the basis $\ket E$ of states that diagonalize the $H$ operator the  classical states are 
\begin{equation}\label{classcohstate}
    \ket{\zeta}_{\mathrm{class}} =\int_0^\infty \psi_\zeta(E) \ket E=\frac{1}{\sqrt{\Gamma(2s+1)}}\qty(\frac{2}{\hbar})^{s+\frac12}\qty(\frac{1-\abs{\zeta}^2}{(1-\zeta)^2})^{s+\frac12}\int_0^\infty \dd E \, E^{s}\exp \left[{-\frac{E}{\hbar}\qty(\frac{1+\zeta}{1-\zeta})}\right]\ket{E}\,,  
\end{equation}
with
\begin{align}
   \zeta = -\tanh s 
\end{align}
The explicit form of the squeezed Glauber coherent state $\ket{\alpha_\zeta}$, with squeezing parameter $c_\zeta$, in the basis of $\ket p$ states that diagonalizes the $P$ operator can be found in \cite{Ciambelli:2025ztm}.

In the large representation-parameter $s$ limit, the entanglement entropy of such classical states gives \cite{Ciambelli:2025ztm}
\begin{equation}\label{eeclassicalstate}
    S^{\mathrm{cl}} \sim  3s + \frac12\ln( s), \quad s\gg1. 
\end{equation}
It is the linear dependence of entanglement entropy on $s$ in the large $s$ limit that makes these states special. As we will soon see, this is directly related to  the area scaling as an entropy in the semiclassical regime. It should be emphasized that the entanglement entropy in \eqref{eeclassicalstate} is not the entanglement entropy of a quantum field on a fixed gravitational background, as introduced in the seminal works \cite{Bombelli:1986rw, Srednicki:1993im}. Rather, it is an entanglement entropy arising from quantum gravity itself.

Before turning to the area–entropy relation, let us first consider the classical aspect of the problem—namely, the expression of corner charges in terms of the metric components for four-dimensional spherically symmetric spacetimes. To connect with the framework developed in our previous works \cite{Ciambelli:2024qgi} and \cite{Ciambelli:2025ztm}, we begin by performing a dimensional reduction from four to two dimensions and then compute the corresponding charges. It is straightforward to verify the one-to-one correspondence between the charges and their algebras in the theory dimensionally reduced to two dimensions and those of spherically symmetric configurations in four dimensions.

We start with the Einstein-Hilbert action in a four-dimensional spacetime $M$ (we work in units where $c=1$)
\begin{equation}\label{eq:einsteinhilbert}
S_{\mathrm{EH}} = \frac{1}{16\pi G}\int_M \sqrt{-g} \,\dd^4 x \,R_4.       
\end{equation}
In the case of spherically symmetric manifold $M$, the metric can be written as
\begin{equation}\label{eq:sphericallysymmetricmetric}
    \dd s^2 = g_{ab}\dd x^a \dd x^b + \rho(x^a)^2 \dd \Omega_S^2,
\end{equation}
where $a,b = 0,1$, $\dd \Omega_S^2$ is the metric on the two sphere and $\rho$ is a scalar field that only depends on the coordinates $(x^0,x^1)$. In order to compute the Ricci scalar of the above metric, we rewrite it
\begin{equation}
    \dd s^2 = \frac{\rho^2}{L^2}\qty(\frac{L^2}{\rho^2} g_{ab} \dd x^a \dd x^b + L^2 \dd \Omega^2_S),
\end{equation}
where $L$ is a length scale needed for dimensional reasons.
The Manifold $M$ is therefore conformally equivalent to a product manifold $\tilde{M}_2\times S^2$
\begin{equation}
    \dd \tilde{s}^2 =  \tilde{g}_{ab}\dd x^a \dd x^b + L^2 \dd \Omega^2_S,
\end{equation}
with 
\begin{equation}\label{eq:tildetogconformaltransformation}
    \tilde{g}_{ab} = \frac{L^2}{\rho^2}g_{ab}.
\end{equation}
The Ricci scalar \( R_4 \) of the spherically symmetric geometry and the Ricci scalar \( \tilde{R}_4 \) of the conformally related geometry are connected through the standard conformal transformation formula:
\begin{equation}
    R_4 = L^2 \left( \rho^{-2} \tilde{R}_4 - 6 \rho^{-3} \tilde{\Box} \rho \right).
\end{equation}
Furthemore, the curvature of the conformal geometry is simply the sum of the curvature of the two-dimensional manifold $\tilde{M_2}$ and the one of the sphere
\begin{equation}
    \tilde{R}_4 = \tilde{R}_2 + \frac{2}{L^2}. 
\end{equation}
This allows us to write the action \eqref{eq:einsteinhilbert} as
\begin{Align}
    S_{\mathrm{EH}} &= \frac{1}{16\pi G}\frac{1}{L^2}\int_M \dd^4 x \sqrt{-\tilde{g}}\, \qty[\rho^2 \qty(\tilde{R}_2 + \frac{2}{L^2}) - 6 \rho \tilde{\Box} \rho]\\
    &= \frac{1}{4 G} \int_{\tilde{M}_2}\dd^2 x \sqrt{-\tilde{g}}\qty[\rho^2\qty(\tilde{R}_2+\frac{2}{L^2}) -6 \rho \tilde{\Box}\rho]\\
    &=\frac{1}{4 G} \int_{\tilde{M}_2}\dd^2 x \sqrt{-\tilde{g}}\qty[\rho^2\qty(\tilde{R}_2+\frac{2}{L^2})+6\partial_a \rho \partial^a \rho] -\frac{3}{4G}\stintb{\partial \tilde{M}_2}{a}\sqrt{-\tilde{g}}\tilde{g}^{ba}\partial_b \qty(\rho^2).\\
\end{Align}

In order to connect the above action to two-dimensional dilaton theories, we define the dimensionless field
\begin{equation}\label{Phidef}
    \Phi = \frac{\rho^2}{4G\hbar}=\frac{\rho^2}{4l_P^2}
\end{equation}
where $l_P=\sqrt{G\hbar}$ is the Planck length. Using this new field, the action reads
\begin{equation}\label{eq:dimreddilatontheory}
    S_{EH} = \hbar\int_{\tilde{M}_2}\dd^2x \sqrt{-\tilde{g}}\qty(\Phi \tilde{R}_2 +2 \frac{\Phi}{L^2}+\frac32 \frac{\partial_a \Phi \partial^a \Phi}{\Phi}) -3\hbar\stintb{\partial \tilde{M}_2}{a}\sqrt{-\tilde{g}}\tilde{g}^{ba}\partial_b \Phi.
\end{equation}
We can now perform a final conformal transformation on the two-dimensional metric
\begin{equation}\label{eq:tildetohatconformaltransformation}
    \tilde{g} = \Phi^{-\frac32}\hat{g}.
\end{equation}
to get the final action 
\begin{equation}\label{eq:dimredaction}
        S_{\mathrm{EH}} = \hbar\int_{\tilde{M}_2} \dd^2 x \sqrt{-\hat{g}}\qty(\Phi \hat{R}_2 + \Phi^{-\frac12}\frac{2}{L^2}) - \frac32 \hbar\stintb{\partial \tilde{M}_2}{a}\sqrt{-\hat{g}}\hat{g}^{ab}\partial_b\Phi.
\end{equation}
This proves the dynamical equivalence between spherically symmetric Einstein-Hilbert theory in four-dimensions and a two-dimensional dilaton gravity model with a $\Phi^{-\frac12}$ potential. In this work, the boundary $\partial \tilde{M}_2$ is a null hypersurface defined by the level set $\Phi=\mathrm{const}$. Its null normal is $k^a=\nabla^a\Phi$, with $k^a k_a=0$. Since the normal $k^a$ enters both the integrand and the surface element $\dd\Sigma_a$, the corresponding boundary contribution vanishes on a null hypersurface. We will thus discard it for the rest of this work.
Finally we write the dictionary between the original four-dimensional metric \eqref{eq:sphericallysymmetricmetric} and the fields in the action \eqref{eq:dimreddilatontheory},
\vspace{-24pt}
\begin{center}
\begin{equation}\label{eq:dictionary}
\boxed{\begin{aligned}
    \Phi &= \frac{\rho^2}{4l_P^2}\\
    \hat g_{ab} &= \frac{\rho\, L^2}{(4l_P^2)^{3/2}}\, g_{ab} 
\end{aligned}%
}   
\end{equation}
\end{center}

 We are now ready to compute the charges of the dimensionally reduced theory. We follow the extended phase space formalism of \cite{Ciambelli:2021nmv}. For convenience, we temporarily work in units where $\hbar=1$ and comment on how to restore $\hbar$ at the end.
 Varying the bulk action yields the following equations of motions
 \begin{align}
    \qty(\hat{\Box} \Phi - \Phi^{-\frac12}\frac{1}{L^2})\hat{g}_{ab} - \hat{\nabla}_{(a} \hat{\nabla}_{b)}\Phi &= 0,\label{eq:eomphi}\\
    \hat{R} &= \frac{\Phi^{-\frac32}}{L^2}\label{eq:eomg}.
 \end{align}
 and the following symplectic potential current
 \begin{equation}
\theta^{\mathrm{bulk}}_{a} = \Phi \hat{\nabla}^b \theta_{ab}- \theta_{ab}\hat{\nabla}^b \Phi,
 \end{equation}
 where 
 \begin{equation}
     \theta_{ab} = \delta \hat{g}_{ab} - \hat{g}_{ab}\hat{g}^{cd}\delta \hat{g}_{cd}.
 \end{equation}
The contraction of the symplectic potential with a field space vector gives
\begin{equation}
I_{\xi}\theta^a_{{\mathrm{bulk}}} \os \hat{\nabla}_b\qty(4 \xi^{[b}\hat{\nabla}^{a]}\Phi + \Phi \hat{\nabla}^{[b}\xi^{a]}) + \frac{3}{L^2} \Phi^{-\frac12}\xi^a. 
\end{equation}
Denoting the bulk Lagrangian density by $\mathcal{L}_\mathrm{bulk}$, the Noether current is given by \cite{Speranza:2022lxr}
\begin{equation}
  J_\xi^a = I_\xi \theta^a_{\mathrm{bulk}} - \xi^a \mathcal{L}_\mathrm{bulk}.   
\end{equation}
We can thus write
\begin{equation}
    J_\xi^{a} \os \hat{\nabla}_b Q^{ab}_\xi,
\end{equation}
with
\begin{equation}
    Q^{ab}_\xi = 4 \xi^{[b}\hat{\nabla}^{a]}\Phi + \Phi \hat{\nabla}^{[b}\xi^{a]}.
\end{equation}
Finally, the Noether charge is given by
\begin{equation}\label{eq:charges}
   H_{\xi} 
   =\frac{\epsilon^{bc}}{2 \sqrt{-\hg}}\qty(\xi^a(4 \hg_{ba}\partial_c \Phi + \Phi \partial_b \hg_{ac}) + \Phi \hg_{ac}\partial_b \xi^a)\eval_{\partial \mathcal{S}},
\end{equation}
where $\epsilon^{ab}$ is the Levi-Civita \textit{symbol} and we used the fact that the connection is torsion free. We note that the charges~\eqref{eq:charges} were obtained after discarding boundary Lagrangian contributions, both in the four-dimensional theory and in its dimensionally reduced version. Although these charges satisfy Hamilton's equations for arbitrary diffeomorphisms within the extended phase space formalism~\cite{Ciambelli:2021nmv}, they are not invariant under the ambiguity transformations of the symplectic structure in the covariant phase space (see, e.g., \cite{Speziale:2025lkm} for a modern review). In the present work, we nevertheless adopt this choice of charges as a convenient arena to illustrate how the area law emerges within the corner proposal. A systematic treatment of boundary terms and the construction of ambiguity-free charges is deferred to future work.

The corner is a single point $ \partial \mathcal{S} = p_S$, we denote the coordinates of the corner point as $\phi(p_S) = \bar{x}^a$. We can then expand the fields and diffeomomrphisms around this point at first order
\begin{align}
    \hg^{ab}(x) &= \hg^{ab}(\bar{x}) + \partial_c \hg^{ab}(\bar{x})\qty(x^c-\bar{x}^c)  \defeq \hg_{(0)}^{ab} + \hg_{(1)c}^{ab}\qty(x^c -\bar{x}^c),\\
   \Phi(x) &= \Phi(\bar{x}) + \partial_a \Phi(\bar{x})\qty(x^a - \bar{x}^a) \defeq \Phi_{(0)} + \Phi^{(1)}_a \qty(x^a - \bar{x}^a),\\
   \xi^{a}(x) &= \xi^a(\bar{x}) + \partial_b \xi^a(\bar{x})\qty(x^b -\bar{x}^b)  \defeq \xi_{(0)}^a + \xi_{(1)b}^a\qty(x^b- \bar{x}^b).\label{eq:diffexpansion}
\end{align}
The charge can then be written
\begin{equation}\label{eq:noetherchargeHxi}
    H_{\xi} = \xi_{(0)}^a t_{a} + \xi_{(1)b}^a N_a^b,
\end{equation}
where, reinstating the explicit value of $\hbar$ we get
\begin{align}
    t_a &= \hbar\frac{\epsilon^{bc}}{2\sqrt{-\hg^{(0)}}}\qty(4 \hg^{(0)}_{ba}\Phi^{(1)}_c + \Phi_{(0)}\hg^{(1)}_{acb}) \\
    N^b_a &= \hbar\frac{\epsilon^{bc}}{2\sqrt{-\hg^{(0)}}}\Phi_{(0)}\hg^{(0)}_{ac},
\end{align}
The seemingly unusual appearance of $\hbar$ in the formulas above stems from the definition of $\Phi$ in \eqref{Phidef}, which explicitly involves $\hbar$.

In terms of the Poisson bracket of the charges $t_a$ generates the commuting normal translations $\mathbb{R}^2$, whereas the traceless $N^b_a$ generates the special linear group $\mathrm{SL}(2,\mathbb{R})$. Using the dictionary \eqref{eq:dictionary}, we can write the charges in terms of the four-dimensional data
\begin{align}
     t_a &= \frac{\epsilon^{bc}}{G\sqrt{-g^{(0)}}}\qty[2 g_{ba}^{(0)}\rho_{(0)}\rho^{(1)}_c + \frac{\rho_{(0)}}{4}\qty(\rho^{(1)}_b g^{(0)}_{ac}+ \rho_{(0)}g^{(1)}_{acb})],\label{eq:classicalcharget}\\
    N^b_a &= \frac{\epsilon^{bc}}{\sqrt{-g^{(0)}}} \frac{\rho^2_{(0)}}{8 G}g^{(0)}_{ac}.\label{eq:classicalchargeN}
\end{align}
Let us also note that the translation charges have dimension of energy while the special linear charge have dimension of energy multiplied by length
\begin{equation}\label{dimensions}
    \qty[t_a] = E, \qquad \qty[N^b_a] = E L.
\end{equation}

\par
After these preliminaries, we can now turn to the main issue we wish to address — namely, the relation between (entanglement) entropy and area. Here we encounter a significant obstacle. Unlike in quantum field theory \cite{Bombelli:1986rw, Srednicki:1993im}, where the geometry is fixed and the notion of area is well defined, in quantum gravity the area itself becomes a quantum gravitational observable. Consequently, comparing the entanglement entropy \eqref{eeclassicalstate} computed for the classical coherent state in eqs.\ \eqref{classcohstate1}, \eqref{classcohstate}, with the horizon area — understood as a function(al) of the spacetime metric — is, in general, meaningful only in the semiclassical limit.\par

To associate the classical value of the area with the quantum representation theory of the QCS, we will make use of the fact that the QCS coherent states \eqref{eq:qcscoherentstate} realize an embedding of the coadjoint orbits in the projective Hilbert space. That is, the parameters $\zeta,\alpha$ can be understood as coordinates on the coadjoint orbits of the QCS. Additionally, the expectation value of the operator in a coherent state provides a quantum moment map \cite{Varrin:2025okc}
\begin{Align}\label{eq:quantummomentmap}
    \mu^{\qty(Q)}: \mathfrak{qcs} &\longrightarrow C^\infty(\Gamma),\\
    V &\longmapsto \mu^{\qty(Q)}(V) = \expval{\hat{V}}{\zeta,\alpha_\zeta}
\end{Align}
where $\Gamma$ is the classical phase space---the coadjoint orbit---, $\hat{V}$ is the quantum operator associated with the Lie algebra generator, and the expectation value is understood as a function on the coadjoint orbits. On the other hand, the Noether charges of the covariant phase space provides a classical moment map
\begin{Align}\label{eq:classicmomentmap}
    \mu^{\qty(C)}: \mathfrak{ecs} &\longrightarrow C^\infty(\Gamma),\\
    V &\longmapsto \mu^{\qty(C)}(V) = H_{\xi_V},
\end{Align}
where $\xi_V$ is a diffeomorphism of the type $\eqref{eq:diffexpansion}$ that is associated with a generator $V\in \mathfrak{ecs}$ in such a way that
\begin{equation}\label{eq:equivariancecondition}
    \qty[\xi_{V_1},\xi_{V_2}]_{\mathrm{Lie}} = \xi_{\qty[V_1,V_2]},
\end{equation}
and $H_{\xi_V}$ is the associated Noether charge \eqref{eq:noetherchargeHxi}. The connection between the quantum and classical data is made by simply identifying those two moment maps $\mu^{(Q)} = \mu^{(C)}$\footnote{More precisely, quantum moment maps live on the $\mathfrak{qcs}$, while classical moment maps live on the $\mathfrak{ecs}$. One can therefore relate the former to the twisted moment maps on the $\mathfrak{ecs}$, and recover the classical theory by sending the central extension to zero (see the companion paper \cite{Varrin:2025okc} for more details). In the present work, however, the central charge cancels in all quantities of interest, so this step is entirely trivial.}.
\par
Let us now move to the description of spherically symmetric static spacetimes (SSS). We consider the SSS in Kruskal-type null coordinates $(U,V,\theta,\phi)$ such that the two-dimensional metric can be written
\begin{equation}\label{eq:metric}
    ds^2 = \frac{f(\rho)}{\kappa^2 UV} \dd U \dd V,
\end{equation}
where $f(\rho)$ is a function whose largest root defines the horizon $f(\rho_h) = 0$, and $\kappa = \frac12 f'(\rho_h)$ is the surface gravity. We take the spacetime subregion as the exterior of the black hole and we define our symplectic structure at a constant (Schwarzschild) time slice. The corner---the intersection between the boundary of the subregion and the constant time slice--- is the bifurcating 2-sphere $U=V=0$, which allows to identify the null coordinates with the coordinates in the expansion \eqref{eq:diffexpansion}. From equations \eqref{eq:classicalchargeN} and \eqref{eq:metric}, it is easy to see that the only non-vanishing $\spl{2}$ charge is the one associated with the boost
\begin{equation}
    \xi_{\mathrm{boost}} = \frac12\qty(V \partial_V - U\partial_U).
\end{equation}
The boost is in the hyperbolic conjugacy class of the special linear transformation and as such, must be associated with an hyperbolic generator in the abstract Lie algebra. Using the equivariance condition \eqref{eq:equivariancecondition} further imposes the identification with the $D$ generator in \eqref{eq:hermitianbasis} $\xi_{\mathrm{boost}} = \xi_D$. We obtain
\begin{equation}
    H_{\xi_D} = -\frac{\rho_h^2}{4 G}.
\end{equation}
This is an expression of the famous relationship between the boost Noether charge and the area of the bifurcating horizon \cite{Wald:1993nt,Jacobson:1993vj,Iyer94}.
\par
To connect with the quantum formalism, we need to find a particular coherent state for which the expectation values of the $\spl{2}$ generators reproduce the structure of the classical charges. That is
\begin{align}
 \expval{D}_{\zeta,\alpha_\zeta} &= H_{\xi_D},\label{eq:areacondition}\\
  \expval{H}_{\zeta,\alpha_\zeta} &= \expval{K}_{\zeta,\alpha_\zeta} = 0. \label{eq:vanishingcondition}
\end{align}
The physical interpretation is the following. Whereas the coadjoint orbits describe the phase space of the classical theory, points on the coadjoint orbit describe particular solutions. The solution to equations \eqref{eq:areacondition} and \eqref{eq:vanishingcondition} in terms of $\alpha,\zeta$ are then answer to the question: Which points in the coadjoint orbits can correspond to the SSS solution? A lengthy but otherwise straightforward computation shows that one may use the second condition,
\eqref{eq:vanishingcondition}, to solve for $\alpha$ (and the constant $c$) as functions of $\zeta$.
Substituting these expressions back into \eqref{eq:areacondition}, the remaining constraint collapses to
the remarkably simple relation
\begin{equation}\label{eq:area_s_identification}
    \frac{4\pi \rho_h^2}{4 l_p^2} =  \pi \qty(4s + 3),
\end{equation}
We see that the classical limit in which the area of the horizon is much bigger than the Planck area corresponds to large values of the representation parameter $s$. This an expected behavior of quantization via coherent states such as Berezin quantization (see the companion paper \cite{Varrin:2025okc} for more details.)
\par
The identification \eqref{eq:area_s_identification} is independent of the particular value of $r = \abs{\zeta}$. In particular, we may choose the classical state~\eqref{classcohstate}. When the system is in this state, the entanglement entropy between the spacetime regions inside and outside the horizon exhibits an area law in the semiclassical regime~\eqref{eeclassicalstate}
\begin{equation}
    S_{\mathrm{cl}} \sim \frac{A}{4\,l_p^2} \;+\; \frac12 \ln\!\Big(\frac{A}{4\,l_p^2}\Big) \;+\; \cdots,
\end{equation}
where the dots denote terms subleading at large area (i.e.\ higher order in the inverse area in Planck units). Moreover, the leading correction is logarithmic as is expected from a quantum gravity theory. As was already pointed out in \cite{Ciambelli:2025ztm}, the coefficient in front of the quantum corrections is universal.

In this letter, we have provided an explicit realization of the Corner Proposal in the setting of four-dimensional spherically symmetric static spacetimes, establishing a direct connection between the representation theory of the QCS introduced and discussed in \cite{Ciambelli:2024qgi,Ciambelli:2025ztm} and the classical geometrical notion of area. By performing the dimensional reduction of the four-dimensional Einstein–Hilbert action to a two-dimensional dilaton gravity model, we derived the classical corner charges and demonstrated that their algebra reproduces the semiclassical limit of the QCS algebra. The use of Perelomov coherent states as ``most classical'' quantum states enabled us to relate expectation values of operators to classical geometric quantities.

We have shown that the entanglement entropy of such classical coherent states obeys the characteristic area scaling in the large-representation-parameter limit, reproducing the Bekenstein–Hawking form
thereby recovering the area law from the representation theory of quantum corners.
This result provides concrete support for the Corner Proposal, which is proven to be a framework capable of reproducing the semiclassical structure of gravity directly from the quantum algebra of boundary observables. Future investigations will aim at understanding the gluing of multiple quantum corners and the role of the central term in encoding subleading quantum corrections to the entropy–area relation.

\paragraph{Acknowledgments}We are deeply grateful to Luca Ciambelli for valuable guidance on the corner formalism, as well as for his foundational contributions to the framework employed in this work. We are additionally indebted to Giulio Neri for stimulating discussions that helped clarify several points. We also thank Rodrigo Andrade e Silva, Michele Arzano, Nicolas Cresto, Michael Imseis, Simon Langenscheidt, Leonardo Sanhueza Mardones, and Simone Speziale for insightful discussions. LV thanks Perimeter Institute for its hospitality. This work is based upon work from COST Action BridgeQG, CA23130, supported by COST (European Cooperation in Science and Technology).

\bibliography{file1}

@article{Ciambelli:2025ztm,
    author = "Ciambelli, Luca and Kowalski-Glikman, Jerzy and Varrin, Ludovic",
    title = "{Entanglement entropy of quantum corners}",
    eprint = "2507.16800",
    archivePrefix = "arXiv",
    primaryClass = "hep-th",
    doi = "10.1007/JHEP02(2026)188",
    journal = "JHEP",
    volume = "02",
    pages = "188",
    year = "2026"
}

@article{Neri:2025fsh,
    author = "Neri, Giulio and Varrin, Ludovic",
    title = "{Orbit method for Quantum Corner Symmetries}",
    eprint = "2507.10683",
    archivePrefix = "arXiv",
    primaryClass = "hep-th",
    month = "7",
    year = "2025"
}

@article{Varrin:2024sxe,
    author = "Varrin, Ludovic",
    title = "{Physical representations of corner symmetries}",
    eprint = "2409.10624",
    archivePrefix = "arXiv",
    primaryClass = "hep-th",
    doi = "10.1103/PhysRevD.111.086003",
    journal = "Phys. Rev. D",
    volume = "111",
    number = "8",
    pages = "086003",
    year = "2025"
}

@article{Ciambelli:2024qgi,
    author = "Ciambelli, Luca and Kowalski-Glikman, Jerzy and Varrin, Ludovic",
    title = "{Quantum corner symmetry: Representations and gluing}",
    eprint = "2406.07101",
    archivePrefix = "arXiv",
    primaryClass = "hep-th",
    doi = "10.1016/j.physletb.2025.139544",
    journal = "Phys. Lett. B",
    volume = "866",
    pages = "139544",
    year = "2025"
}

@article{Ciambelli:2021nmv,
	archiveprefix = {arXiv},
	author = {Ciambelli, Luca and Leigh, Robert G. and Pai, Pin-Chun},
	doi = {10.1103/PhysRevLett.128.171302},
	eprint = {2111.13181},
	journal = {Phys. Rev. Lett.},
	primaryclass = {hep-th},
	title = {{Embeddings and Integrable Charges for Extended Corner Symmetry}},
	volume = {128},
	year = {2022}}

@article{Ciambelli:2021vnn,
	archiveprefix = {arXiv},
	author = {Ciambelli, Luca and Leigh, Robert G.},
	doi = {10.1103/PhysRevD.104.046005},
	eprint = {2104.07643},
	journal = {Phys. Rev. D},
	number = {4},
	pages = {046005},
	primaryclass = {hep-th},
	title = {{Isolated surfaces and symmetries of gravity}},
	volume = {104},
	year = {2021}}

@article{Donnelly:2016auv,
	archiveprefix = {arXiv},
	author = {Donnelly, William and Freidel, Laurent},
	doi = {10.1007/JHEP09(2016)102},
	eprint = {1601.04744},
	journal = {JHEP},
	pages = {102},
	primaryclass = {hep-th},
	title = {{Local subsystems in gauge theory and gravity}},
	volume = {09},
	year = {2016}}

@article{Ciambelli:2022vot,
    author = "Ciambelli, Luca",
    title = "{From Asymptotic Symmetries to the Corner Proposal}",
    eprint = "2212.13644",
    archivePrefix = "arXiv",
    primaryClass = "hep-th",
    doi = "10.22323/1.435.0002",
    journal = "PoS",
    volume = "Modave2022",
    pages = "002",
    year = "2023"
}

@article{APerelomov_1977,
doi = "10.1070/PU1977v020n09ABEH005459",
year = "1977",
volume = "20",
number = "9",
pages = "703",
author = "A M Perelomov",
title = "Generalized coherent states and some of their applications",
journal = "Soviet Physics Uspekhi"
}

@article{Ciambelli:2023bmn,
    author = "Ciambelli, Luca and D'Alise, Alessandra and D'Esposito, Vittorio and Dj{}ordj{}evic, Du\v{s}an and Fern\'andez-Silvestre, Diego and Varrin, Ludovic",
    title = "{Cornering quantum gravity}",
    eprint = "2307.08460",
    archivePrefix = "arXiv",
    primaryClass = "hep-th",
    doi = "10.22323/1.440.0010",
    journal = "PoS",
    volume = "QG-MMSchools",
    pages = "010",
    year = "2024"
}

@article{Jacobson:1993vj,
    author = "Jacobson, Ted and Kang, Gungwon and Myers, Robert C.",
    title = "{On black hole entropy}",
    eprint = "gr-qc/9312023",
    archivePrefix = "arXiv",
    reportNumber = "MCGILL-93-22, NSF-ITP-93-152, UMDGR-94-75",
    doi = "10.1103/PhysRevD.49.6587",
    journal = "Phys. Rev. D",
    volume = "49",
    pages = "6587--6598",
    year = "1994"
}

@article{Wald:1999wa,
    author = "Wald, Robert M. and Zoupas, Andreas",
    title = "{A General definition of 'conserved quantities' in general relativity and other theories of gravity}",
    eprint = "gr-qc/9911095",
    archivePrefix = "arXiv",
    doi = "10.1103/PhysRevD.61.084027",
    journal = "Phys. Rev. D",
    volume = "61",
    pages = "084027",
    year = "2000"
}

@article{Wald:1993nt,
    author = "Wald, Robert M.",
    title = "{Black hole entropy is the Noether charge}",
    eprint = "gr-qc/9307038",
    archivePrefix = "arXiv",
    reportNumber = "EFI-93-42",
    doi = "10.1103/PhysRevD.48.R3427",
    journal = "Phys. Rev. D",
    volume = "48",
    number = "8",
    pages = "R3427--R3431",
    year = "1993"
}

@article{Speranza:2022lxr,
    author = "Speranza, Antony J.",
    title = "{Ambiguity resolution for integrable gravitational charges}",
    eprint = "2202.00133",
    archivePrefix = "arXiv",
    primaryClass = "hep-th",
    doi = "10.1007/JHEP07(2022)029",
    journal = "JHEP",
    volume = "07",
    pages = "029",
    year = "2022"
}

@article{Ciambelli:2022cfr,
    author = "Ciambelli, Luca and Leigh, Robert G.",
    title = "{Universal corner symmetry and the orbit method for gravity}",
    eprint = "2207.06441",
    archivePrefix = "arXiv",
    primaryClass = "hep-th",
    doi = "10.1016/j.nuclphysb.2022.116053",
    journal = "Nucl. Phys. B",
    volume = "986",
    pages = "116053",
    year = "2023"
}

@article{Speziale:2025lkm,
    author = "Speziale, Simone",
    title = "{A short introduction to boundary symmetries}",
    eprint = "2512.16810",
    archivePrefix = "arXiv",
    primaryClass = "hep-th",
    month = "12",
    year = "2025"
}

@misc{Iyer94,
  title = {Some properties of the Noether charge and a proposal for dynamical black hole entropy},
  author = {Iyer, Vivek and Wald, Robert M.},
  journal = {Phys. Rev. D},
  volume = {50},
  issue = {2},
  pages = {846--864},
  numpages = {0},
  year = {1994},
  month = {Jul},
  publisher = {American Physical Society},
  doi = {10.1103/PhysRevD.50.846},
  url = {https://link.aps.org/doi/10.1103/PhysRevD.50.846},
howpublished = {}
}

@article{Wigner:1939cj,
    author = "Wigner, Eugene P.",
    editor = "Kim, Y. S. and Zachary, W. W.",
    title = "{On Unitary Representations of the Inhomogeneous Lorentz Group}",
    doi = "10.2307/1968551",
    journal = "Annals Math.",
    volume = "40",
    pages = "149--204",
    year = "1939"
}

@article{Bombelli:1986rw,
    author = "Bombelli, Luca and Koul, Rabinder K. and Lee, Joohan and Sorkin, Rafael D.",
    title = "{A Quantum Source of Entropy for Black Holes}",
    reportNumber = "PRINT-86-0371 (SYRACUSE)",
    doi = "10.1103/PhysRevD.34.373",
    journal = "Phys. Rev. D",
    volume = "34",
    pages = "373--383",
    year = "1986"
}

@article{Srednicki:1993im,
    author = "Srednicki, Mark",
    title = "{Entropy and area}",
    eprint = "hep-th/9303048",
    archivePrefix = "arXiv",
    reportNumber = "LBL-33754, CFPA-93-02",
    doi = "10.1103/PhysRevLett.71.666",
    journal = "Phys. Rev. Lett.",
    volume = "71",
    pages = "666--669",
    year = "1993"
}

@book{Weinberg:1995mt,
    author = "Weinberg, Steven",
    title = "{The Quantum theory of fields. Vol. 1: Foundations}",
    doi = "10.1017/CBO9781139644167",
    isbn = "978-0-521-67053-1, 978-0-511-25204-4",
    publisher = "Cambridge University Press",
    month = "6",
    year = "2005"
}

@article{Rosas-Ortiz:2018jum,
    author = "Rosas-Ortiz, Oscar",
    title = "{Coherent and Squeezed States: Introductory Review of Basic Notions, Properties, and Generalizations}",
    eprint = "1812.07523",
    archivePrefix = "arXiv",
    primaryClass = "quant-ph",
    doi = "10.1007/978-3-030-20087-9_7",
    year = "2019"
}

@article{Yuen:1976vy,
    author = "Yuen, H. P.",
    title = "{Two photon coherent states of the radiation field}",
    doi = "10.1103/PhysRevA.13.2226",
    journal = "Phys. Rev. A",
    volume = "13",
    pages = "2226--2243",
    year = "1976"
}

@article{Stoler:1969tq,
    author = "Stoler, David",
    title = "{Equivalence classes of minimum uncertainty packets}",
    doi = "10.1103/PhysRevD.1.3217",
    journal = "Phys. Rev. D",
    volume = "1",
    pages = "3217--3219",
    year = "1970"
}

@article{Glauber:1963tx,
    author = "Glauber, Roy J.",
    title = "{Coherent and incoherent states of the radiation field}",
    doi = "10.1103/PhysRev.131.2766",
    journal = "Phys. Rev.",
    volume = "131",
    pages = "2766--2788",
    year = "1963"
}

@misc{Varrin:2025okc,
  author       = {Varrin, Ludovic},
  title        = {Semiclassical limit of quantum gravity on corners},
  howpublished = {Accepted for publication in Phys. Rev. D, \href{https://doi.org/10.1103/kkny-kkt4}{doi:10.1103/kkny-kkt4}},
  year         = {2026},
  eprint       = {2510.25843},
  archivePrefix = {arXiv},
  primaryClass = {hep-th}
}

\end{document}